\documentclass[lettersize,journal]{IEEEtran} 
\usepackage{cite}
\usepackage{mathrsfs}
\usepackage{amsthm}
\usepackage[cmex10]{amsmath}
\usepackage{mathtools}
\usepackage{array}
\usepackage{eqparbox}
\usepackage{bm}
\newtheorem{myDef}{Definition}
\usepackage[table,dvipsnames]{xcolor}
\usepackage{multicol,booktabs,tabularx}

\usepackage{cases}
\usepackage{algorithm}
\usepackage{algpseudocode}

\usepackage{graphicx}
\usepackage{subfigure}
\usepackage{epstopdf}
\usepackage{epsfig}
\usepackage{amssymb}
\usepackage{array}
\usepackage{multirow}
\newtheorem{theorem}{Theorem}

\begin{document}

\title{On the Performance of an Integrated Communication and Localization System: an Analytical Framework}

\author{Yuan Gao, Haonan Hu, Jiliang Zhang, \textit{Senior Member, IEEE}, Yanliang Jin, Shugong Xu, \textit{Fellow, IEEE} and Xiaoli Chu, \textit{Senior Member, IEEE}
\thanks{This paper is supported by the Innovation Program of Shanghai MunicipaScience and Technology Commission under Grant 22511103202.}
\thanks{Yuan Gao, Yanliang Jin and Shugong Xu are with the School of Communication and Information Engineering, Shanghai University, P.R.C, email: gaoyuansie@shu.edu.cn, jinyanliang@staff.shu.edu.cn and shugong@shu.edu.cn.}
\thanks{Haonan Hu. is with the School of Communication and Information Engineering, Chongqing University of Posts and Telecommunications, Chongqing, P.R.C, e-mail: huhn@cqupt.edu.cn.}
\thanks{Jiliang Zhang is with the College of Information Science and Engineering, Northeastern University, Shenyang, P.R.C, email: zhangjiliang1@mail.neu.edu.cn.}
\thanks{Xiaoli Chu is with the Department of Electronic and Electrical Engineering, The University of Sheffield, UK, e-mail: x.chu@sheffield.ac.uk.}
}
\maketitle

\begin{abstract}
Quantifying the performance bound of an integrated
localization and communication (ILAC) system and the trade-off between communication and localization performance is critical. In this letter, we consider an ILAC system that can perform communication and localization via time-domain or frequency-domain resource allocation. We develop an analytical framework to derive the closed-form expression of the capacity loss versus localization Cramer-Rao lower bound (CRB) loss via time-domain and frequency-domain resource allocation. Simulation results validate the analytical model and demonstrate that frequency-domain resource allocation is preferable in scenarios with a smaller number of antennas at the next generation nodeB (gNB) and a larger distance between user equipment (UE) and gNB, while time-domain resource allocation is preferable in scenarios with a larger number of antennas and smaller distance between UE and the gNB.

\end{abstract}


\section{Introduction}
Integrated sensing and communication (ISAC) that performs communication and sensing simultaneously by leveraging the same spectrum resource and hardware has been considered a key technology of 6G \cite{zhang2021enabling}. Sensing is expected to trigger various novel applications, such as smart cities, smart transportation systems and industrial Internet of Things (IoT). For the communication industries and operators, localization has been the most urgent and desirable sensing function, and has attracted wide attention \cite{del2017survey}.  

Resource allocation has been studied for integrated localization and communication (ILAC) without revealing the fundamental performance bound and trade-off quantitatively. The ILAC performance trade-off was first investigated by allocating dedicated time-domain resources for communication and localization. respectively \cite{kumar2018trade,destino2017trade}. On top of the time-domain resource sharing, the localization accuracy and data rate are further explored by optimizing the beam width for communication and localization, respectively \cite{ghatak2020beamwidth}. A waveform optimization scheme was developed to maximize the mutual information of joint communication and localization \cite{yuan2020spatio}. A beamforming optimization scheme was proposed to maximize the communication capacity while guaranteeing localization accuracy \cite{kwon2021joint}. Joint time-frequency-spatial domain resource sharing enables a balance between data rate and localization error in \cite{koirala2019localization,bao2022resource}, while perfect CSI is assumed, which is impractical in real ILAC systems. In addition, all the above works are based on simulation without exposing the fundamental performance limits or trade-off between communication and localization in the time-frequency domain analytically, which fails to predict the ILAC performance in a tractable way and provides quite limited insights for real ILAC system design.

To overcome the limitations of the existing research, we propose an analytical framework for an ILAC system considering the channel estimation overhead and error, where spectrum resource is allocated for dedicated communication  and localization in the time-domain and frequency-domain. The simulation results validate that our analytical model fairly approximates the ILAC performance in time-domain and frequency-domain resource allocation with great predictability. Using this analytical model, we further reveal the ILAC performance bound and the trade-off between communication and localization performance, which provide insights for practical ILAC system design.
\section{System Model}
\begin{figure}[htbp]
\centering\includegraphics[width=0.48\textwidth]{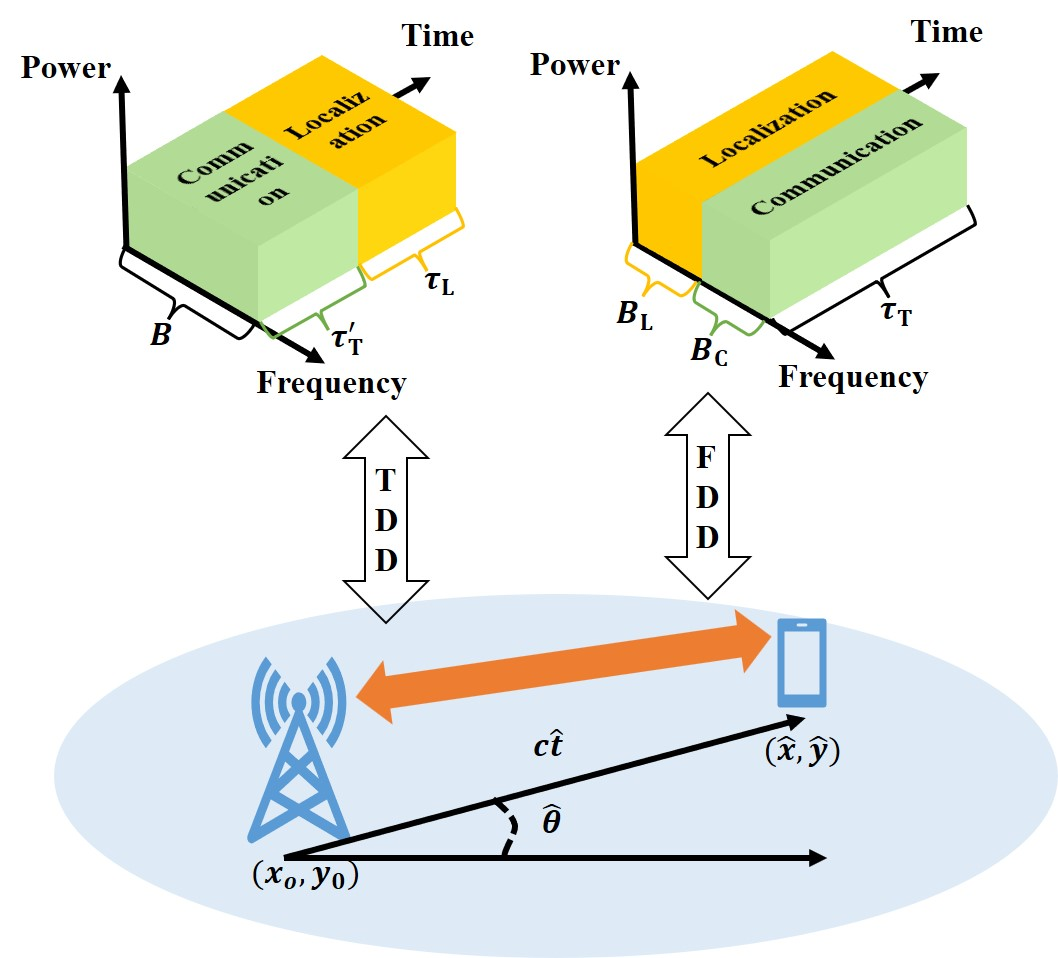}
\caption{System model of ILAC network, resource allocation in time- and frequency-domain}
\label{system_model}
\end{figure}

As shown in Fig. \ref{system_model}, we consider an ILAC 5G NR system, where a 5G gNB equipped with a uniform linear array (ULA) of $N_\mathrm{T}$ antennas (antenna space of $d_\mathrm{A}$) transmits downlink communication signals to the user equipment (UE) and transmits positioning reference signal for localization. 

Without loss of generality, the frame structure of the time-frequency resource in 5G is divided into communication and localization blocks for simplicity. The time-domain resource allocation for ILAC is illustrated in the top left of Fig. \ref{system_model}, where a total number of $\tau_\mathrm{T}$ symbols is divided into the communication block of $\tau'_\mathrm{T}$ and the localization block with $\tau_\mathrm{L}$. For the frequency-domain resource allocation in the top right of Fig. \ref{system_model}, the spectrum resource of total bandwidth of $B$ is divided into the communication block of $B_\mathrm{C}$ and the localization block with $B_\mathrm{L}$. The communication block consists of channel estimation consisting of $\tau_\mathrm{P}$ symbols and data transmission consisting of $\tau'_\mathrm{L}-\tau_\mathrm{P}$ symbols. 

We consider this network to work in a downlink-only
ILAC mode while uplink pilot is used for channel estimation.
Hence, the capacity of such network using maximum ratio is
calculated as \cite{marzetta2016fundamentals}:
\begin{equation}\label{capacity}
C(B_\mathrm{C},\tau'_\mathrm{T} )=B\mathrm{C}\frac{\tau'_\mathrm{T}-\tau_\mathrm{P}}{\tau_\mathrm{T}}\mathrm{In}\left(1+\frac{N_\mathrm{T}\rho_\mathrm{dl}\nu }{\rho_\mathrm{dl}\beta  } \right ),
\end{equation}where $\beta$ is the large-scale fading coefficient and $\nu=(\tau_\mathrm{P}\rho_\mathrm{ul}\beta ^2)/(1+\tau_\mathrm{P}\rho_\mathrm{ul}\beta)$ is the mean square of the channel
estimation \cite{marzetta2016fundamentals}. $\rho_\mathrm{ul}=P_\mathrm{UE}G_\mathrm{gNB}G_\mathrm{UE}/N$ and $\rho_\mathrm{dl}=P_\mathrm{gNB}G_\mathrm{gNB}G_\mathrm{UE}/N$ are the nominal uplink and downlink
signal to noise ratio (SNR), respectively. $N$ is the white
noise. This framework can be easily extended to the uplinkonly
ILAC mode or the time-division UL-DL ILAC mode by
exploiting the uplink capacity also given in \cite{marzetta2016fundamentals}.
\section{ILAC}
In the considered ILAC system in Fig. \ref{system_model}, each UE is served by the gNB that offers the largest received signal strength, while performing 2D-localization using the same gNB as an anchor with known location $(x_0,y_0)$. The estimated localization $(\widehat{x},\widehat{y})$ of the UE calculated as:
\begin{small}\begin{subequations}\label{xy_localization}                                                                                                      
\begin{align} &\widehat{x}=x_0+\mathrm{cos}(\widehat{\theta})\mathrm{c}\widehat{t}\\    
&\widehat{y}=y_0+\mathrm{sin}(\widehat{\theta})\mathrm{c}\widehat{t}
\end{align}
\end{subequations}\end{small}where $\widehat{t}$ is the estimated time of arrival (ToA), $\widehat{\theta}$ is the angle of arrival (AoA), $\theta$ is the angle between the UE and the orientation of the ULA, and $c$ is the speed of light. 

The localization CRB based on joint ToA and AoA estimation using  Eq. (\ref{xy_localization}) is given in \cite{richards2014fundamentals} as: 
\begin{small}
\begin{equation}\label{CRB_L}
CRB_\mathrm{L}(B_\mathrm{L},\tau_\mathrm{L})=\mathrm{c}^2\widehat{t}^2CRB_{\widehat{\theta}}(\tau_\mathrm{L})+\mathrm{c}^2CRB_{\widehat{t}}(B_\mathrm{L},\tau_\mathrm{L})
\end{equation}\end{small}where $CRB_{\widehat{\theta}}(\tau_\mathrm{L})$ and $CRB_{\widehat{t}}(B_\mathrm{L},\tau_\mathrm{L})$ are the CRB of ToA and AoA estimation \cite{richards2014fundamentals}, respectively, given as:

\begin{small}\begin{equation}\label{CRB_theta}
CRB_{\widehat{\theta}}(\tau_\mathrm{L})=\frac{3\lambda^2}{4\pi^2d_\mathrm{A}^2\gamma\mathrm{cos}^2\theta N_\mathrm{T}(N_\mathrm{T}-1)(2N_\mathrm{T}-1)\tau_\mathrm{L}}
\end{equation}\end{small}
\begin{small}\begin{equation}\label{CRB_time}
CRB_{\widehat{t}}(B_\mathrm{L},\tau_\mathrm{L})=\frac{3}{8\pi^2B_\mathrm{L}^2(1+\varsigma)\gamma N_\mathrm{T}\tau_\mathrm{L}}
\end{equation}\end{small}where $\gamma$ is the SNR, $\varsigma$ is the coefficient dependent on the waveform, and $\lambda$ is the wavelength of the spectrum.


\subsection{Time-Domain Resource Allocation}

In the time-domain resource allocation ILAC illustrates in the top left of Fig. \ref{system_model}, we develop the following performance analysis framework. 
\begin{myDef}The system capacity loss attributed to allocating $\tau_\mathrm{L}$ symbols for localization is calculated as: 
\begin{small}\begin{equation}
L^\mathrm{t}_\mathrm{C}=C(B,\tau_\mathrm{T})-C(B,\tau'_\mathrm{T})
\end{equation}\end{small}where $\tau'_\mathrm{T}=\tau_\mathrm{T}-\tau_\mathrm{L}$.
\end{myDef}
\begin{myDef}The system localization CRB loss attributed to allocating $\tau'_\mathrm{T}$ symbols for communication is calculated as: 
\begin{small}\begin{equation}\label{L_L}
L^\mathrm{t}_\mathrm{L}=\frac{CRB_L(B,\tau_\mathrm{L})}{CRB_L(B,\tau_\mathrm{T})}=\frac{\tau_\mathrm{T}}{\tau_\mathrm{L}}
\end{equation}\end{small}
\end{myDef}

\begin{theorem}
\label{Loss_time}
In time domain resource allocation, the relationship between localization CRB loss $L^\mathrm{t}_\mathrm{L}$ and capacity loss $L^\mathrm{t}_\mathrm{C}$ is expressed as:
\begin{small}\begin{equation}\label{Loss_LC_time}
L^\mathrm{t}_\mathrm{L}=\tau_\mathrm{T}\left(\frac{\mathrm{In}\left(\frac{\alpha}{1+\delta}\right)}{\sqrt{\frac{N_\mathrm{T}\varepsilon \mathrm{In}\left(\frac{\alpha}{1+\delta}\right)}{\alpha}}+\sqrt{\frac{N_\mathrm{T}\varepsilon \mathrm{In}\left(\frac{\alpha}{1+\delta}\right)}{\alpha}+\frac{L^\mathrm{t}_\mathrm{C}\tau_\mathrm{T}\mathrm{In}\left(\frac{\alpha}{1+\delta}\right)}{B}}}\right)^2,
\end{equation}\end{small}where $\alpha=1+(N_\mathrm{T}+1)\rho_\mathrm{dl} \beta$, $\varepsilon=\rho_\mathrm{dl}/\rho_\mathrm{dl}$ and $\delta=\rho_\mathrm{ul} \beta$.
\end{theorem}
\begin{proof}Proof of \textbf{Theorem} \ref{Loss_time} refers to Appendix \ref{SecondAppendix}. \end{proof}

\begin{figure*} \centering  
\subfigure[8 antennas scenario.] { 
\label{CL_bound_8}     
\includegraphics[width=0.95\columnwidth]{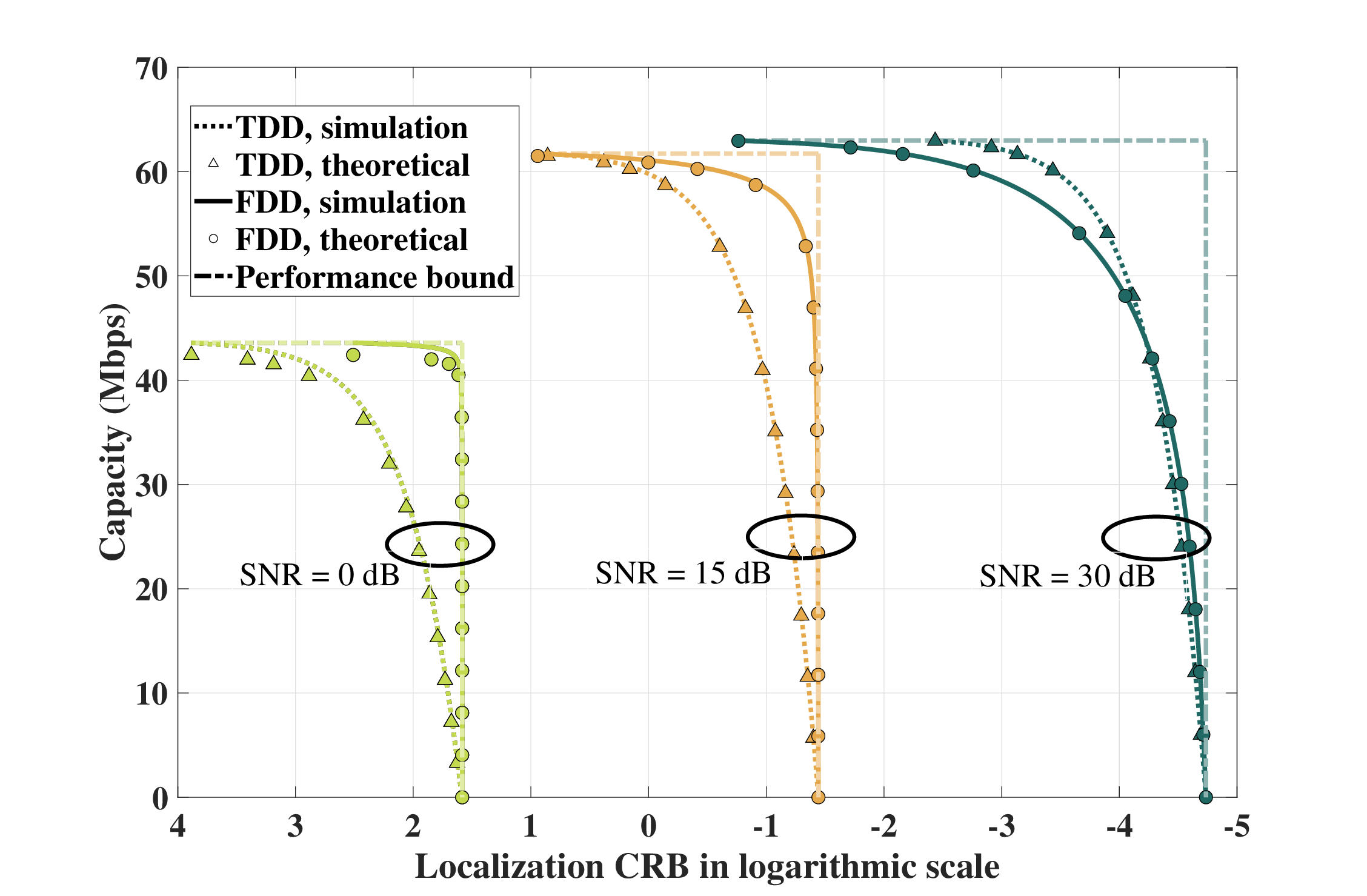}     
}    
\subfigure[32 antennas scenario.] { 
\label{CL_bound_32}     
\includegraphics[width=0.95\columnwidth]{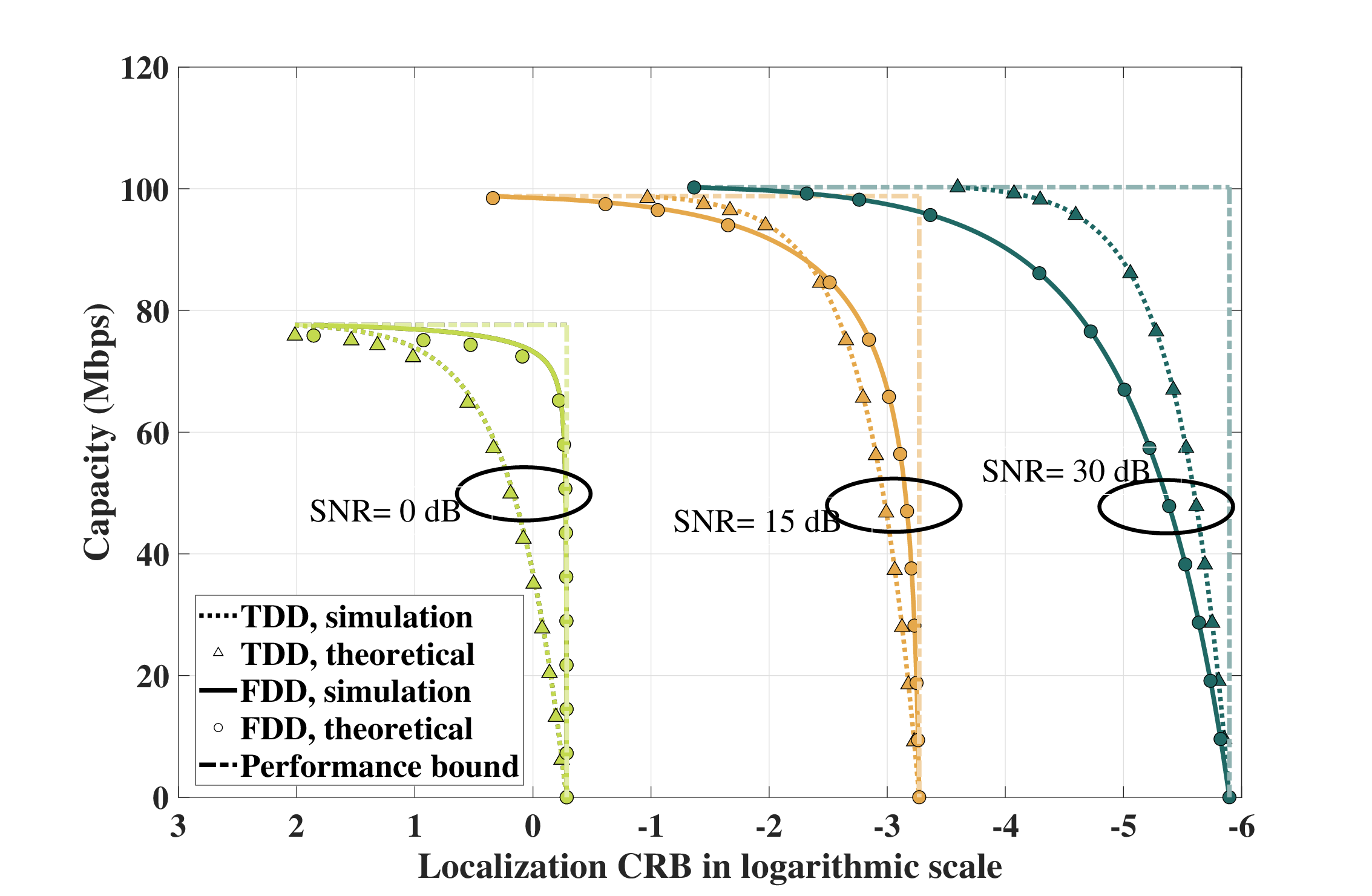}     
}   
\caption{Capacity versus CRB in frequency (solid line) and time (dotted line) domain. Time-domain and frequency-domain resource allocation are denoted as TDD and FDD, respectively.}     
\label{CL_bound}     
\end{figure*}
\begin{figure*} \centering    
\subfigure[8 antennas scenario.] {
 \label{Loss_CL_8}     
\includegraphics[width=0.95\columnwidth]{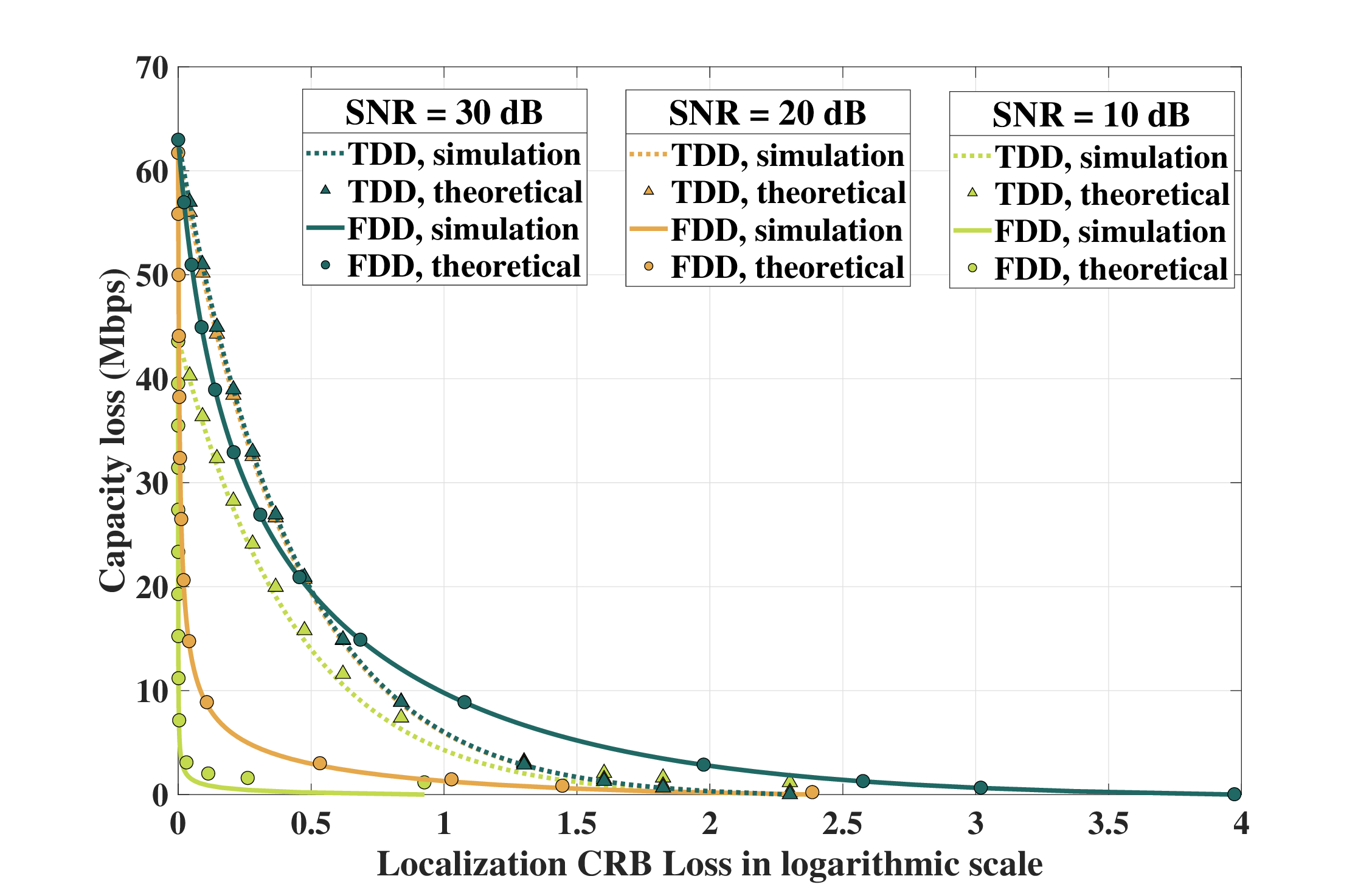}  
}     
\subfigure[32 antenna scenario.] {
 \label{Loss_CL}     
\includegraphics[width=0.95\columnwidth]{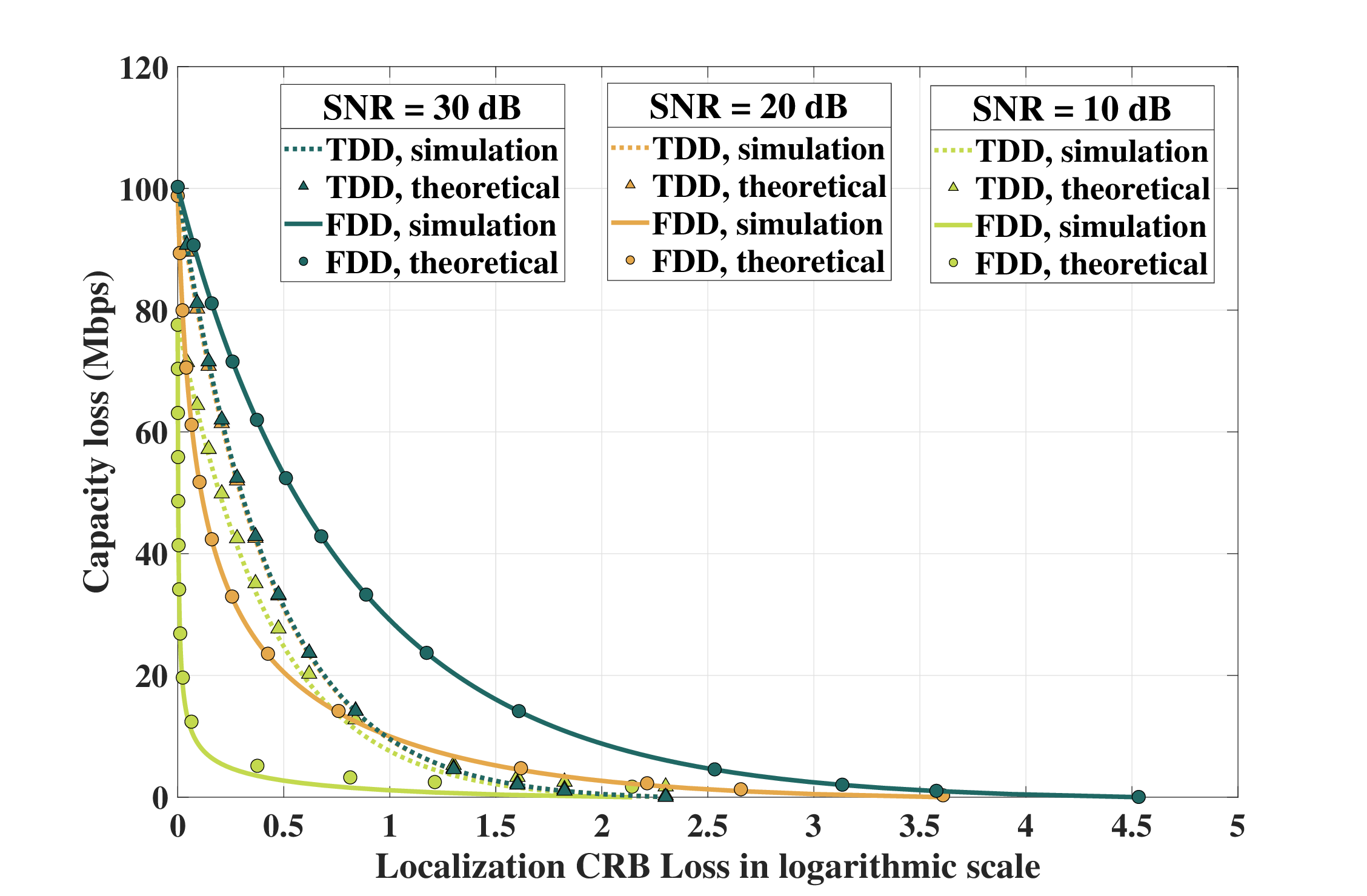}  
}
\caption{Capacity loss versus CRB loss in frequency (solid line) and time (dotted line) domain. Time-domain and frequency-domain resource allocation are denoted as TDD and FDD, respectively.}     
\label{Loss_CL}     
\end{figure*} 

\subsection{Frequency-Domain Resource Allocation}
In the frequency-domain resource allocation ILAC illustrates in the top right of Fig. \ref{system_model}, we develop the following performance analysis framework. 

\begin{myDef}
The system capacity loss attributed to allocating bandwidth $B_\mathrm{L}$ for localization is calculated as: 
\begin{small}\begin{equation}
L^\mathrm{f}_\mathrm{C}=C(B,\tau_\mathrm{T})-C(B_\mathrm{C},\tau_\mathrm{T})
\end{equation}\end{small}where $B_\mathrm{C}=B-B_\mathrm{L}$.
\end{myDef}
\begin{myDef}
The system localization CRB loss attributed to allocating bandwidth $B_\mathrm{C}$ communication is calculated as: 
\begin{small}\begin{equation}\label{L_L_f}
L^\mathrm{f}_\mathrm{L}=\frac{CRB(B_\mathrm{L},\tau_\mathrm{T})}{CRB(B,\tau_\mathrm{T})}
\end{equation}\end{small}
\end{myDef}

\begin{theorem}
\label{Loss_frequency}
In frequency domain resource allocation, the relationship between localization CRB loss $L^\mathrm{f}_\mathrm{L}$ and capacity loss $L^\mathrm{f}_\mathrm{C}$ is expressed as:
\begin{small}
\begin{equation}\label{F_CL_loss}
L^\mathrm{f}_\mathrm{L}=\frac{ \frac{3t^2\lambda^2}{d_\mathrm{A}^2\mathrm{cos}^2\theta(N_\mathrm{T}-1)(2N_\mathrm{T}-1)}+\frac{12}{2B_\mathrm{L}^2(1+\varsigma)}}{\frac{3t^2\lambda^2}{d_\mathrm{A}^2\mathrm{cos}^2\theta(N_\mathrm{T}-1)(2N_\mathrm{T}-1)}+\frac{12}{2B^2(1+\varsigma)}}
\end{equation}\end{small}where $B_\mathrm{L}$ is calculated as:
\begin{small}\begin{equation}
B_\mathrm{L}=\frac{L^\mathrm{f}_\mathrm{C}\tau_\mathrm{T}}{\tau_\mathrm{T}\mathrm{In}\left(\frac{\alpha}{1+\delta}\right)-2\sqrt{\frac{\tau_\mathrm{T}N_\mathrm{T}\varepsilon\mathrm{In}\left(\frac{\alpha}{1+\delta}\right) }{\alpha }}+\frac{N_\mathrm{T}\varepsilon }{\alpha}}
\end{equation}
\end{small}$\alpha=1+(N_\mathrm{T}+1)\rho_\mathrm{dl} \beta$, $\varepsilon=\rho_\mathrm{dl}/\rho_\mathrm{dl}$ and $\delta=\rho_\mathrm{ul} \beta$.
\end{theorem}
 \begin{proof}Proof of \textbf{Theorem} \ref{Loss_frequency} refers to Appendix \ref{ThirdAppendix}. \end{proof}

\section{Simulation Results}
\subsection{Simulation Settings}

The ILAC system is operating on the 2.6 GHz spectrum with a maximum bandwidth of 20 MHz and a sub-carrier bandwidth of 180 KHz. The coherence time of this system contains 200 symbols. The minimum resource element of this system is 180
KHz × 1 symbol, which can be allocated for communication or localization. $\zeta=1$ is designed for optimal localization performance in the position reference signal. We consider the resource block for communication and localization to be continuous. Free-space path loss model is adopted to calculate the large-scale fading coefficient $\beta$. We consider a single-antenna UE and multiple-antenna omnidirectional gNB
(implying that $\theta = 0$) with antenna numbers of 8 and 32, and antenna space of $d_\mathrm{A} = \lambda /2$. The antenna gain of gNB $G_\mathrm{gNB}$
and UE $G_\mathrm{UE}$ are calculated using the antenna beamforming
gain in [12]. $P_\mathrm{gNB}$ = 13dBm, $P_\mathrm{gUE}$ = 13dBm.

\subsection{Simulation Results Analysis}
In Fig. 2, the trade-off between the capacity and the localization CRB in the time-domain and frequency-domain are demonstrated in scenario with 8 (Fig. 2(a)) and 32 antenna (Fig. 2(b)), respectively. The maximum capacity and minimum localization CRB are plotted in horizontal and vertical lines, respectively. The general insight is that the frequency-domain allocation outperforms the time-domain allocation in low SNR and/or smaller number of antennas scenarios, such as $SNR = 10dB$ and $20dB$ Fig. 2(a), and $SNR = 10dB$ in Fig. 2(b). When increasing the antenna number, time-domain allocation tends to become effective, as demonstrated by comparing the $SNR = 30dB$ results of Fig. 2(a) and Fig. 2(b). Timedomain allocation outperforms frequency-domain allocation in $SNR = 30dB$ scenario in Fig. 2(b).

Due to the fundamental trade-off between capacity and localization CRB, it makes more sense to reveal the effectiveness of sacrificing capacity for localization gain, which is demonstrated in Fig. 3. A significant observation is that the capacity loss versus localization CRB loss achieved by time-domain resource allocation is much less affected by SNR than that achieved by frequency-domain resource allocation. Nevertheless, according to both Eq. (11) and Fig. 3, the capacity loss versus localization CRB loss achieved by frequency-domain resource allocation is much affectively by SNR. Specifically, a higher level gain of localization CRB can be achieved by sacrificing a certain level of capacity if SNR is smaller, i.e., larger distance between the gNB and UE. In addition, as indicated in Fig. 2, Fig. 3 also demonstrated that the number of antennas affects significantly the effectiveness of time-domain and frequency-domain resource allocation to acquire localization CRB gain by sacrificing a certain level of capacity. With a smaller number of antennas, frequency-domain resource allocation outperforms time-domain resource allocation.

The above simulation results can provide further insights into achieving a better localization performance by choosing a resource allocation approach (time-domain or frequency-domain) concerning the number of antennas and SNR, i.e., distance between gNB and UE. In a nutshell, it is preferable to allocate frequency-domain resources for localization in scenarios with a smaller number of antennas and smaller SNR (larger distance between the gNB and UE), i.e., outdoor scenarios. While time-domain resource allocation is preferable in indoor scenarios with a larger number of antennas and larger SNR (smaller distance between the gNB and UE).
\section{Conclusion}
In this article, we derived the closed-form performance bound of the communication and localization in an ILAC system. We also reveal the analytical relationship between the loss of capacity and the loss of localization CRB, and the fundamental effect of time-domain and frequency-domain resource allocation in terms of achieving the localization accuracy gain by sacrificing a certain level of capacity in various scenarios with different sizes. Such insights will provide guidance to exploit the radio resource more effectively to achieve communication and localization simultaneously in ILAC systems. We will extend our ILAC model into multiple-gNB and multiple-UE scenarios for a more comprehensive analysis.
\appendix

\subsection{Proof of \textbf{Theorem} \ref{Loss_time}}
\label{SecondAppendix}
\begin{proof}
To analytically analyse the capacity concerning the number of symbols $\tau_\mathrm{P}$ allocated for channel estimation, we apply the inequality \cite{sheng2018low}:
\begin{small}\begin{equation}\label{inequation}
\mathrm{In}(1+x)\ge \mathrm{In}(1+\overline{x})+\frac{\overline{x}}{1+\overline{x}}(1-\frac{\overline{x}}{x}),
\end{equation}\end{small}to transform Eq. (\ref{capacity}) as: 
\begin{small}\begin{subequations}
\begin{align}
&C(B_\mathrm{C},\tau'_\mathrm{T})\cong \nonumber\\
&\frac{B_\mathrm{C}}{\tau_\mathrm{T}}\left(\tau'_\mathrm{T}\mathrm{In}\left(\frac{\alpha}{1+\delta}\right)-\tau'_\mathrm{T}\frac{N_\mathrm{T} \varepsilon }{\alpha\tau_\mathrm{P}}-\tau_\mathrm{P}\mathrm{In}\left(\frac{\alpha}{1+\delta}\right)+\frac{N_\mathrm{T}\varepsilon  }{\alpha}\right)\tag{14}.
\end{align}\end{subequations}\end{small}where $\alpha=1+(N_\mathrm{T}+1)\rho_\mathrm{dl} \beta$, $\varepsilon=\rho_\mathrm{dl}/\rho_\mathrm{dl}$ and $\delta=\rho_\mathrm{ul} \beta$.

The optimal $\tau_\mathrm{P}$ that maximizes $C$ can be easily calculated as: 
\begin{small}\begin{equation}
\widetilde{\tau}_\mathrm{P}=\sqrt{\frac{\tau'_\mathrm{T}N_\mathrm{T} }{\alpha \mathrm{In}\left(\frac{\alpha}{1+\delta}\right)}}.
\end{equation}\end{small}
By substituting $\tau_\mathrm{P}$ with $\widetilde{\tau}_\mathrm{P}$, we have the optimal capacity for ${C}(B,\tau_\mathrm{T})$ and ${C}(B,\tau'_\mathrm{T})$ as $\widetilde{C}(B_\mathrm{LB},\tau_\mathrm{T})$ and $\widetilde{C}(B_\mathrm{LB},\tau'_\mathrm{T})$, respectively. For time domain resource allocation, the loss of capacity $L^\mathrm{t}_\mathrm{C}$ is calculated as:
\begin{small} 
\begin{subequations}
\begin{align}\label{L_C_T}
&L^\mathrm{t}_\mathrm{C}=\widetilde{C}(B,\tau_\mathrm{T})-\widetilde{C}(B,\tau'_\mathrm{T})=\nonumber\\
&\frac{B}{\tau_\mathrm{T}}\left(\tau_\mathrm{L}\mathrm{In}\left(\frac{\alpha}{1+\delta}\right)-2\sqrt{\frac{\tau_\mathrm{L}N_\mathrm{T}\mathrm{In}\left(\frac{\alpha}{1+\delta}\right) }{\alpha }}\right).\tag{16}
\end{align}
\end{subequations}
\end{small}By solving Eq. (\ref{L_C_T}) we have:
\begin{small}
\begin{equation}\label{tau_L}
\tau_\mathrm{L}=\left(\frac{\sqrt{\frac{N_\mathrm{T}\mathrm{In}\left(\frac{\alpha}{1+\delta}\right)}{\alpha}}+\sqrt{\frac{N_\mathrm{T}\mathrm{In}\left(\frac{\alpha}{1+\delta}\right)}{\alpha}+\frac{L_\mathrm{C}\tau_\mathrm{T}\mathrm{In}\left(\frac{\alpha}{1+\delta}\right)}{B}}}{\mathrm{In}\left(\frac{\alpha}{1+\delta}\right)}\right)^2.
\end{equation}
\end{small}
By integrating Eq. (\ref{tau_L}) into Eq. (\ref{L_L}), we have the capacity loss and localization CRB loss in Eq. (\ref{Loss_LC_time}). This closed the proof of \textbf{Theorem} \ref{Loss_time}.
\end{proof}

\subsection{Proof of \textbf{Theorem} \ref{Loss_frequency}}
\label{ThirdAppendix}
\begin{proof}For frequency domain resource allocation, the loss of capacity $L_C$ is calculated as:
\begin{small} 
\begin{subequations}
\begin{align}\label{L_C_F}
&L^\mathrm{f}_\mathrm{C}=\widetilde{C}(B,\tau_\mathrm{T})-\widetilde{C}(B_\mathrm{C},\tau_\mathrm{T})=\nonumber\\
&\frac{B-B_\mathrm{C}}{\tau_\mathrm{T}}\left(\tau_\mathrm{T}\mathrm{In}\left(\frac{\alpha}{1+\delta}\right)-2\sqrt{\frac{\tau_\mathrm{T}N_\mathrm{T}\mathrm{In}\left(\frac{\alpha}{1+\delta}\right) }{\alpha }}+\frac{N_\mathrm{T} }{\alpha}\right)\tag{20}.
\end{align}
\end{subequations}
\end{small}By solving Eq. (\ref{L_C_T}) we have:
\begin{small}\begin{equation}\label{B_L}
B_\mathrm{L}=\frac{L^\mathrm{f}_\mathrm{C}\tau_\mathrm{T}}{\tau_\mathrm{T}\mathrm{In}\left(\frac{\alpha}{1+\delta}\right)-2\sqrt{\frac{\tau_\mathrm{T}N_\mathrm{T}\mathrm{In}\left(\frac{\alpha}{1+\delta}\right) }{\alpha }}+\frac{N_\mathrm{T} }{\alpha}}.
\end{equation}\end{small}
By integrating Eq. (\ref{B_L}) into Eq. (\ref{F_CL_loss}), we have the capacity loss and localization CRB loss in Eq. (\ref{F_CL_loss}). This closed the proof of \textbf{Theorem} \ref{Loss_frequency}.
\end{proof}
\bibliographystyle{IEEEtran}
\bibliography{Myreference}
\end{document}